# EdgeLaaS: Edge Learning as a Service for Knowledge-Centric Connected Healthcare

Gaolei Li, Guangquan Xu, Arun Kumar Sangaiah, Jun Wu, and Jianhua Li

## Abstract

By introducing networking technologies and services into healthcare infrastructures (e.g., multimodal sensors and smart devices) that are deployed to supervise a person's health condition, the traditional healthcare system is being revolutionized toward knowledge-centric connected healthcare (KCCH), where persons will take their own responsibility for their healthcare in a knowledge-centric way. Due to the volume, velocity, and variety of healthcare supervision data generated by these healthcare infrastructures, an urgent and strategic issue is how to efficiently process a person's healthcare supervision data with the right knowledge of the right guardians (e.g., relatives, nurses, and doctors) at the right time. To solve this issue, the naming and routing criterion of medical knowledge is studied. With this offloaded medical knowledge, we propose an edge learning as a service (EdgeLaaS) framework for KCCH to locally process health supervision data. In this framework, edge learning nodes can help the patient choose better advice from the right guardians in real time when some emergencies occur. Two application cases: 1) fast self-help and 2) mobile help pre-calling are studied. Performance evaluations demonstrate the superiority of KCCH and EdgeLaaS, respectively.

## Introduction

The World Health Organization (WHO) reports that the best way to strengthen our health systems is no longer solely outsourcing it to professional doctors but taking our own responsibility for it [1]. In next-generation connected healthcare, the Internet of things (IoT) will play an important role [2, 3]. Therefore, the motivation for developing next generation connected healthcare is to give us a global and in-depth health view by associating islands of healthcare IoT data that are separated [4]. For example, the heart rate sensed by a smart watch can be associated with the latest blood testing results at a hospital to infer the health condition at home. In next generation connected healthcare, it is imperative to build person-scale connected healthcare (PCH) systems and homescale connected healthcare (HCH) systems to provide real-time healthcare responses. PCH often consists of several multi-modal sensors or smart devices, which can aggregate and analyze health supervision data in real time. In PCH, the person often specifies a smart device as a local data center to store and process her/his health supervision data. Several PCH systems can be connected and form an HCH system. Both HCH systems and PCH systems support healthcare services, acting as edge nodes of connected heathcare.

After collecting the health supervision data, an urgent and strategic issue is how to securely and efficiently process these data with correct knowledge of the right guardians, such as relatives, nurses, and doctors, at the right time [5, 6]. For the persons in connected healthcare, they will have three different methods to analyze their data:

1. Local processing
2. Outsourcing to hospitals
3. Third party

As stated by the WHO, local processing may be more and more popular because health supervision data are sensitive. However, in order to enable local processing, professional knowledge should be open for each person. It means that, different from the traditional data-centric connected healthcare (DCCH) system where persons need to upload their data to hospitals or third parties, knowledge-centric connected healthcare (KCCH) aims to help people retrieve medical knowledge from hospitals or third parties. In this article, medical knowledge is a unified appellation of various healthcare decisions such as expert diagnosis for some illnesses, nursing skills for different patients, and medication strategies.

Considering the speciality of next generation connected healthcare on knowledge retrieval and distribution, in this article, we first propose to construct a KCCH system. We then study the naming and routing criteria of medical knowledge. To locally mine knowledge from the healthcare supervision data as soon as possible, we propose an edge learning as a service (EdgeLaaS) framework for KCCH. In this framework, edge learning nodes can help the patient choose better advice from the right guardians in real time when some emergencies occur. Figure 1 shows the comparison between DCCH and KCCH.

The remainder of this article is structured as follows. We comprehensively review the roadmap from DCCH to KCCH. We describe the proposed EdgeLaaS framework for KCCH. We introduce two application cases of EdgeLaaS in KCCH. The experimental results are provided to demonstrate EdgeLaaS's superiority.

*Gaolei Li, Jun Wu (corresponding author), and Jianhua Li are with Shanghai Jiao Tong University and the Shanghai Key Laboratory of Integrated Administration Technologies for Information Security; Guangquan Xu (corresponding author) is with the Big Data School at Qingdao Huanghai University and Tianjin University; Arun Kumar Sangaiah is with Vellore Institute of Technology.*

## Road Map to KCCH

The deployment of Internet of Things (IoT) devices in healthcare systems brings exponential data growth. To enable low-latency healthcare applications, cloud/fog computing is envisioned as a key enabling technology in DCCH. However, due to security and privacy, these enabling technologies are not widely deployed in real healthcare systems. As the recent development of artificial intelligence, edge devices are generally equipped with sufficient resources for high-performance computing. Considering that medical knowledge can be open to each user, knowledge generating and routing will be dominated in next generation connected healthcare. Inspired by this idea, we propose KCCH to facilitate the interaction between intelligent IoT edges.

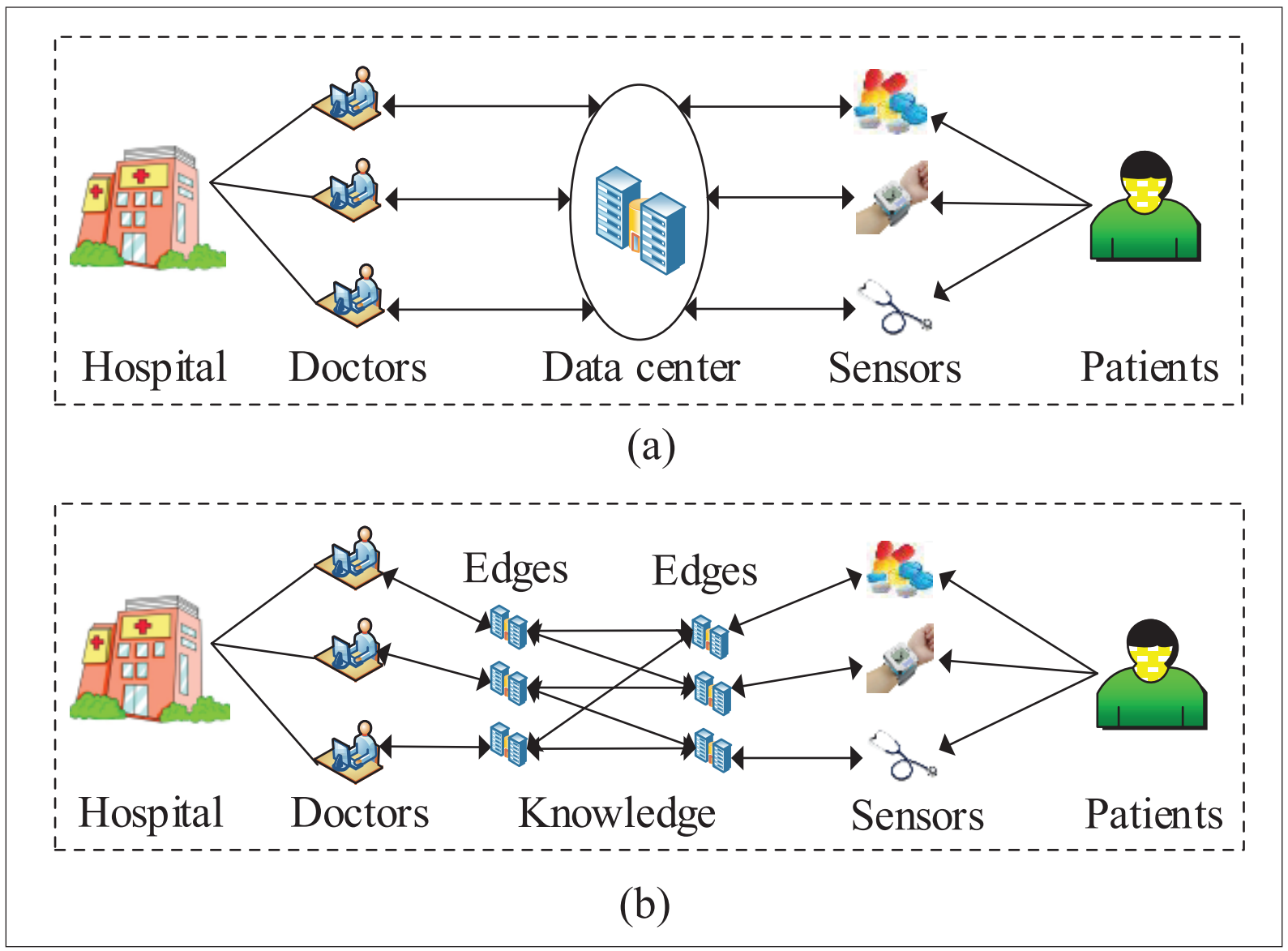

FIGURE 1. The architectures of data-centric connected healthcare (DCCH) and knowledge-centric connected healthcare (KCCH).

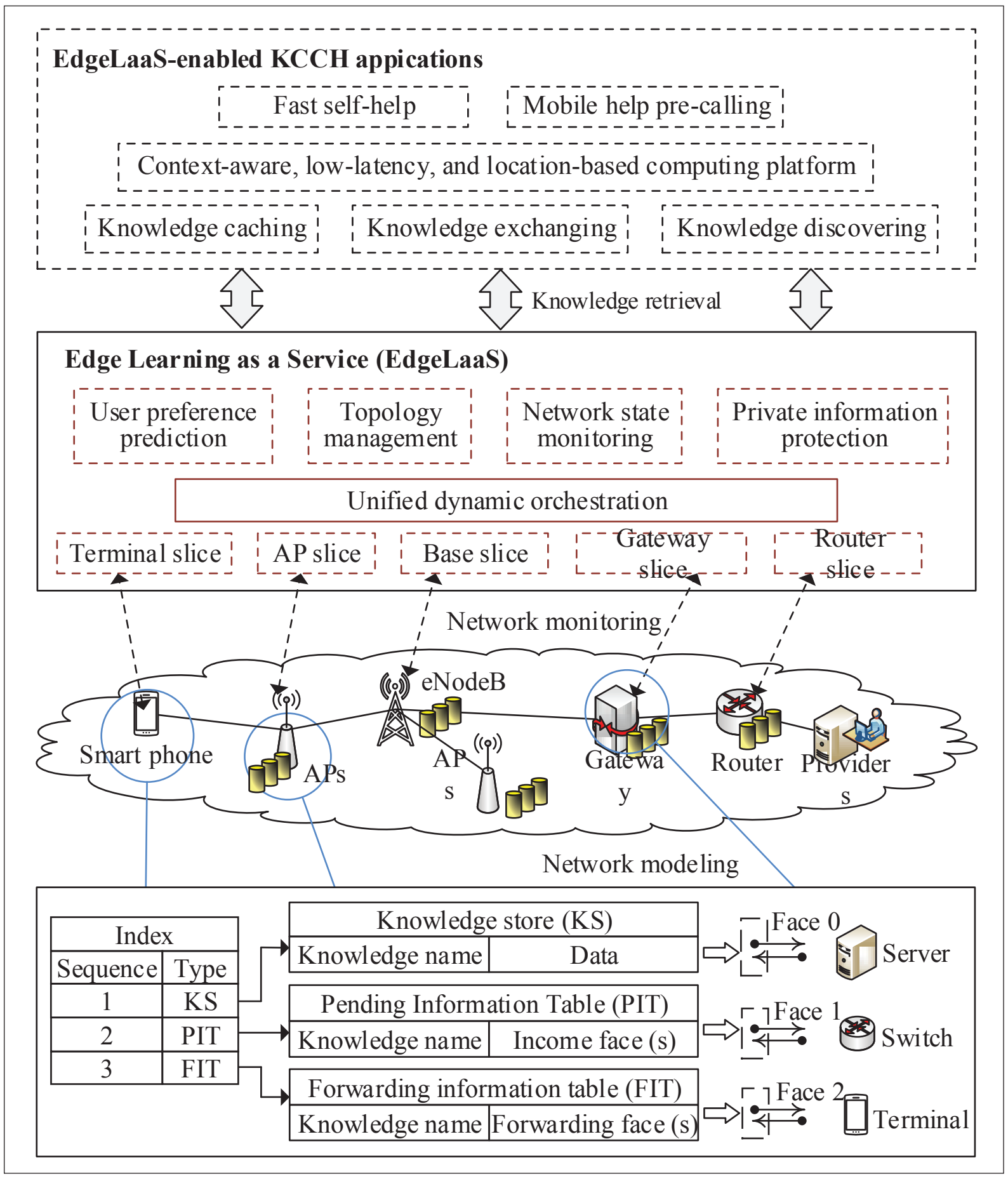

FIGURE 2. Implementation structure of the proposed KCCH.

### Data-Centric Connected Healthcare

In recent decades, cloud computing has evolved to be applied in many scenarios, where data are required to be aggregated into the data center for storing and processing. Cloud computing was a popular solution for hospitals to build a pervasive patient health monitoring system [7, 8]. However, it was hard to circumvent the drawbacks of cloud on privacy preserving and long communication latency [9]. Recently, fog computing with features (e.g., geographically distributed, delay-sensitive, and context-aware) has been considered to deal with the increasing healthcare IoT data [10]. Physical healthcare IoT devices can be virtualized as multiple virtual nodes and monitored by the hierarchical fog nodes, where some lightweight data processing tasks are being done in real time. With such network virtualization and hierarchical computing, the quality and security of healthcare IoT services are improved significantly [11].

In DCCH, fog computing is responsible to run the functions of data preprocessing and supports delay-sensitive applications, while the in-depth data analysis is outsourced to hospital clouds. However, traditional fog computing has limited functions such as data cleansing, data integration, data specification and data transformation, so it is only good at preprocessing known illness cases [12, 13]. To support various applications in next generation connected heathcare, fog computing still needs improvement. The concept of the edge learning platform has been developed by some vendors (e.g., Google Edge TPU and Raspberry Pi+Tensorflow).

### Knowledge-Centric Connected Healthcare

Different from traditional healthcare, where the medical knowledge can only be attained by doctors through long-time learning, edge devices can also grasp some medical knowledge using artificial intelligence. This enforces DCCH to be transferred as KCCH, where a patient can retrieve medical knowledge to provide real-time healthcare IoT services. To construct an efficient KCCH, this article explores the naming and routing criterion of medical knowledge and ultimately proposes the EdgeLaaS framework for KCCH. Figure 2 shows the full view of KCCH.

**Knowledge Generating and Storing:** To resist knowledge poisoning in KCCH, knowledge generating must be checked by professional doctors, and it is independent of common users. Knowledge storing still depends on cloud computing and fog computing. In the traditional cloud paradigm, the learning tasks are often distributed into one or multiple virtual machines in the data center. While the data center has almost inexhaustible computing resources that can provide guarantees to large-scale learning tasks, the long distance between users and data center makes it infeasible to support low-latency healthcare IoT applications. Compared to the cloud/fog computing for DCCH, the cloud/fog computing for KCCH is more scalable and flexible. The cloud/fog of DCCH stores large-scale healthcare data, while the cloud/fog of KCCH stores medical knowledge. It can be imagined that the maintenance cost of DCCH is larger than KCCH because data updating is far more frequent than knowledge updating.

**Knowledge Naming and Routing:** Knowledge naming is the fist step toward KCCH. There are two types of naming methods:

1. Unified naming
2. Patient-oriented naming

Unified naming denotes the medical knowledge with a unique identity (ID) that can be generated by a universally unique identifier (UUID). Unified naming is beneficial to knowledge management but may not be friendly to patients because it requires patients to know the name of the medical knowledge they want.

Patient-oriented naming relies on the attributes of healthcare supervision data. For example, a group of prescriptions acting as medical knowledge can be retrieved by imputing: *Patient_no./B.gender.age/S.B_t.H_r.B_P/P.Chest.Throat.Waist/A.Hair.Pollen*. Patient-oriented naming is very scalable for future extension, and it enables fuzzy searching (i.e., keyword-based searching).

Under an emergency scenario, a person cares more about when she/he can achieve the right medical knowledge but less about where the right medical knowledge comes from. Since the features of medical knowledge retrieval and distribution is similar to content retrieval and distribution, knowledge naming and routing can be built as illustrated in Fig. 2. KCCH defines three elements:

1. Knowledge store (KS)
2. Pending interest table (PIT)
3. Forwarding information table (FIT)

For each element, the knowledge name is defined for fast matching.

When a router receives a request with the above format (10./B.1.20/S.37.78.110/P.0.1.0/A.1.0), it will find the corresponding prescription from the KS. If the corresponding prescription is found in the KS, it will directly send it back to the requester. Otherwise, if there no corresponding prescription in the KS, this request will be forwarded to the next router and then added into the PIT. The matching process between request and prescription is formulated as an edge learning service, which is introduced in the next section.

**Strengths of KCCH:** The strengths of KCCH contain three divisions.

1. Agile response. Different from DCCH, KCCH enables a person to achieve healthcare services by actively retrieving medical knowledge. With an efficient naming and routing mechanism, agile response to a patient's demands can be provided. Moreover, medical knowledge providers can also push some medical information to the medical knowledge caching servers at the network edges (close to the users) to provide an accurate healthcare response.
2. Less congestion. Real-time healthcare monitoring based on IoT is very weak to congestion. A person may aggregate a lot of healthcare supervision data using her/his smart portable devices (e.g., mobile phones, tablets, and smart bands). When she/he wants to know her/his health state, all of these data should be submitted to the hospital cloud for deep analysis in DCCH. It is common to see much congestion due to the limited bandwidth resources and dynamic wireless channel states. KCCH reduces such congestion by distributing medical knowledge into network edges.
3. Mobility support. Medical knowledge is maintained by network edges in a distributed way. Due to the openness of medical knowledge, knowledge exchange among these network edges is freeing. Mobile patients can retrieve medical knowledge from any domain quickly without complicated synchronization or authentication.
4. Privacy protection. KCCH is also beneficial to privacy protection because the healthcare data is not outsourced to third parties but processed at the local area.

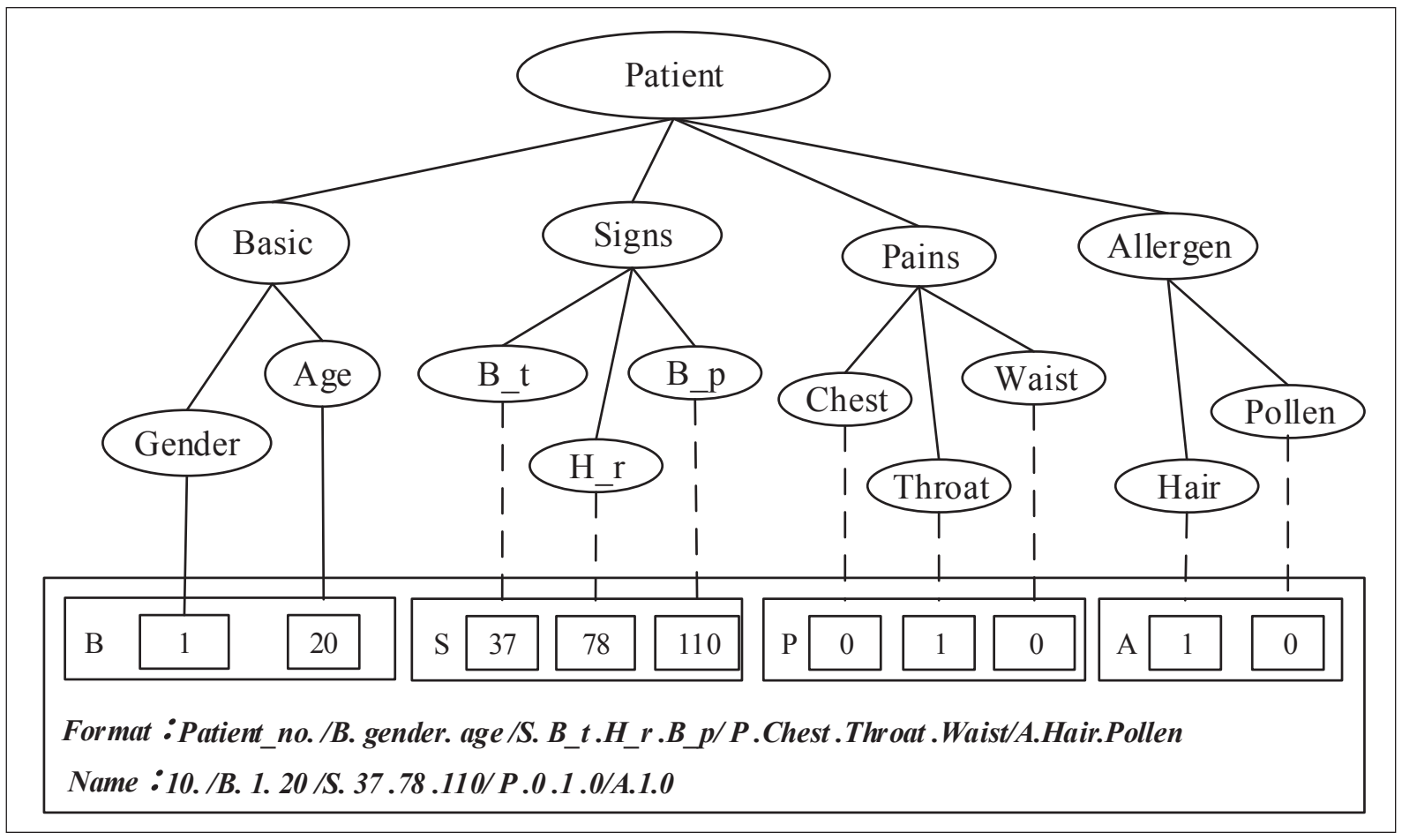

FIGURE 3. Knowledge naming in KCCH.

## EdgeLaaS: Edge Learning as a Service

During the routing process of KCCH, a router needs to match the request with the prescription. However, the knowledge name in a request message is often incomplete. If some key information is missing, it will take a matching mistake, which may bring serious harm. The EdgeLaaS framework for KCCH is proposed to improve the hit rate of matching. EdgeLaaS has the following three features:

1. It is open for each user to access medical knowledge.
2. There is no healthcare center, so each node can provide healthcare IoT services.
3. EdgeLaaS is based on reinforcement learning, which needs to maximize the rewards.

### Stakeholders of EdgeLaaS

Knowledge routers close to users are the service objects of EdgeLaas. We treat these knowledge routers as edge nodes to cache knowledge. The direct stakeholder of EdgeLaaS is the knowledge router because the EdgeLaas can reduce the matching latency by enabling pre-training. The indirect stakeholder of EdgeLaaS is the requester because the EdgeLaaS will predict the knowledge she/he wants. Besides, the knowledge providers can also achieve benefits from EdgeLaaS indirectly because they can dispatch their resources to optimize their knowledge for popular symptoms.

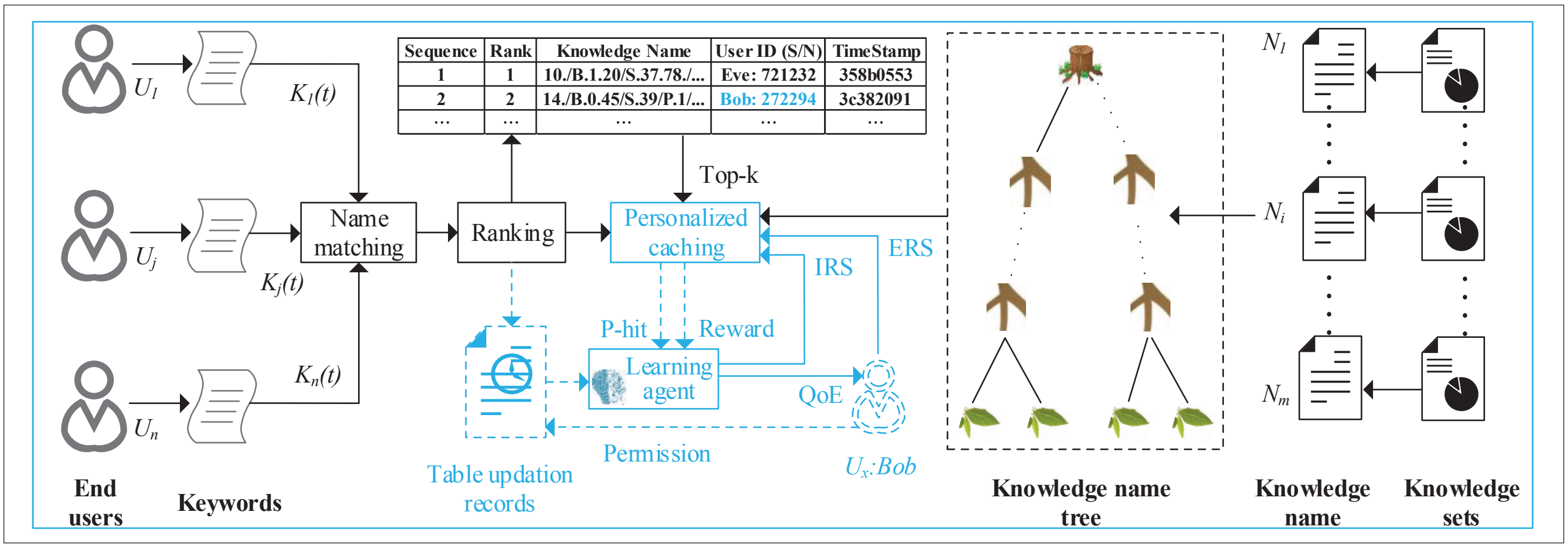

FIGURE 4. Fast self-help with EdgeLaaS-based knowledge pre-caching.

### Modeling of EdgeLaaS

EdgeLaaS is an online learning system based on reinforcement learning. The basic models of EdgeLaaS consist of request sets, knowledge sets, state transferring rules, rewards, and observability range. Full observability is configured in EdgeLaaS so that the local states are completely open for the authorized edge node.

For each request at time slot $t$, the alternative knowledge has different costs (e.g., energy consumption, cache space, and bandwidth occupation) and different rewards (e.g., the number of simultaneously satisfied requests). We calculate the cost and rewards based on multi-armed bandit (MAB) [14].

### Service Orchestration

To orchestrate the edge learning services of KCCH dynamically, we allocate the learning tasks on virtual fog rather than edge entities. The available approaches to enforce distributed learning contain model parallelism and data parallelism. In the data center, model parallelism claims more frequent interactions between slaves than data parallelism, so data parallelism presents lower operation costs. However, different from the simple task distribution among virtual machines in the data center, placement of the learning tasks on virtual fogs is challenging because it is uneconomical to slice the data into multiple geographically distributed fogs. In this article, EdgeLaaS will hierarchically configure different processing models on different fog domains and place the learning tasks through model parallelism. Meanwhile, in the same fog domain, all fog entities (e.g., end devices, access points, and base stations) cooperatively complete the placed learning tasks by data parallelism.

## Application Cases of EdgeLaaS

To clarify the advantages of EdgeLaaS, two application cases are studied:

1. Fast self-help
2. Mobile help calling

Due to limitations of doctors, especially in disasters, it is almost impossible to provide sufficient healthcare services for each patient. Fast self-help is an important application that next-generation connected healthcare should support. Compared to smart decision support systems [15], the proposed EdgeLaaS enables ultra-low-latency self-help services by shifting medical knowledge from the remote data center close to patients. Mobile help pre-calling focuses on off-site treatment of patients. For difficult illnesses in mountain areas, sending patients to hospitals will consume a lot of precious time, during which EdgeLaaS in KCCH can assist by pre-calling doctors for remote diagnosis. In addition, to reduce the cost of the EdgeLaas framework, we present the unified orchestration of edge resources, where we can minimize the iteration number in EdgeLaaS with given knowledge retrieval and network usage.

### Fast Self-Help

Fast self-help is the most important application case of EdgeLaaS. To enable fast self-help, predicting the required self-help knowledge for users is a trend. As shown in Fig. 4, EdgeLaaS predicts the self-help knowledge in which users are interested according to the local users' knowledge retrieval records. Edge nodes can securely aggregate the local users' knowledge retrieval records and then selectively exploit fog computing to optimize self-knowledge caching hit rate and the loss function of EdgeLaaS. As the external reinforcement signals in EdgeLaaS, the desensitized retrieval records of other users that are open to the healthcare stakeholders can also be utilized to optimize fast self-help services during the progress of end-to-end knowledge retrieval and transmission. With the predicted self-help knowledge, KCCH can enforce the acceleration service of knowledge transmission by caching the top-$k$ self-help knowledge at network edges.

The top-$k$ self-help knowledge caching is independent of the knowledge requesters and knowledge providers. However, fast self-help requires binding the user identifiers with knowledge names. In the IP network, the available binding approaches contain the following two different types. The first one is adding a special segment into the IP header, and the second one is configuring some explains on the packet payload. The existing approaches in the IP network cannot be directly utilized to enable fast self-help knowl-edge delivery. First, adding a special segment into a knowledge name may incur heavy name resolution and a series of additional processing costs. Second, configuring the explains on a data packet of KCCH is invalid because the knowledge transmission closely relies on the interest packet.

It is common to see that there are $n$ types of end users (denoted as $\{U_1, U_2, ..., U_j, U_n\}$) that want to obtain self-help knowledge from the Internet over KCCH. Usually, end users need to enter their keywords to retrieve the target self-help knowledge. To improve the service quality (e.g., hit rate of targeted knowledge), a translating function should be provided to precisely associate these entered keywords and translate them into the knowledge names that requesters want.

EdgeLaaS extends the ranking function of the top-$k$ caching strategy by saving the table update records and putting these records into learning agents in virtual fogs. To explain the principle of fast self-help, we specify a self-help knowledge requester $U_x : Bob$ with a special identifier. On the basis of the proposed EdgeLasS framework, the principle of fast self-help consists of the following steps:

Step 1: $U_x : Bob$ sends a message to a learning agent and gives it permission to access table update records of $U_x : Bob$.

Step 2: The learning agent aggregates the network states. Then it comprehensively analyzes the last reward (including the initial reward), network states, and $U_x : Bob$'s records to make a new caching strategy.

Step 3: Repeat step 2 up to service termination, relying on a periodic learning plan or an event-driven learning plan.

## Mobile Help Pre-Calling

Due to the patient-oriented nature of KCCH, the mobility management of patients needs to be well handled. If a patient moves to a new place and calls for help through a new access point (AP), the new AP needs to refresh the knowledge caching strategy to fulfil the patient's demands. However, this refreshing process may pose additional overhead including computation consumption, service quality jitter, and so on. Moreover, with the rapid advancement of device-to-device (D2D) communication in mobile wireless networks, the communication transmission between these mobile devices has gained increasing popularity. The location change of a mobile device acting as a middle hop in the routing path of D2D communication will greatly impact the user's help call. Moreover, the mobility of nearest knowledge sources will incur many invalid requests and massive redirections.

Thus, we exploit EdgeLaaS to enhance the mobility management of both patients and helpers that can provide low knowledge caching refresh time, low service interruption, and low helper discovery overhead. Figure 5 shows the workflow of mobile help pre-calling. For convenience, we use $U_R$ to denote the mobile help requester and $-U_R$ to denote all help requesters in the access domain excluding $U_R$. Correspondingly, if $U_R$ moves into a new access domain, its label will be updated as $U_R'$, and the currently connected help requesters in the new domain are denoted with $-U_R'$.

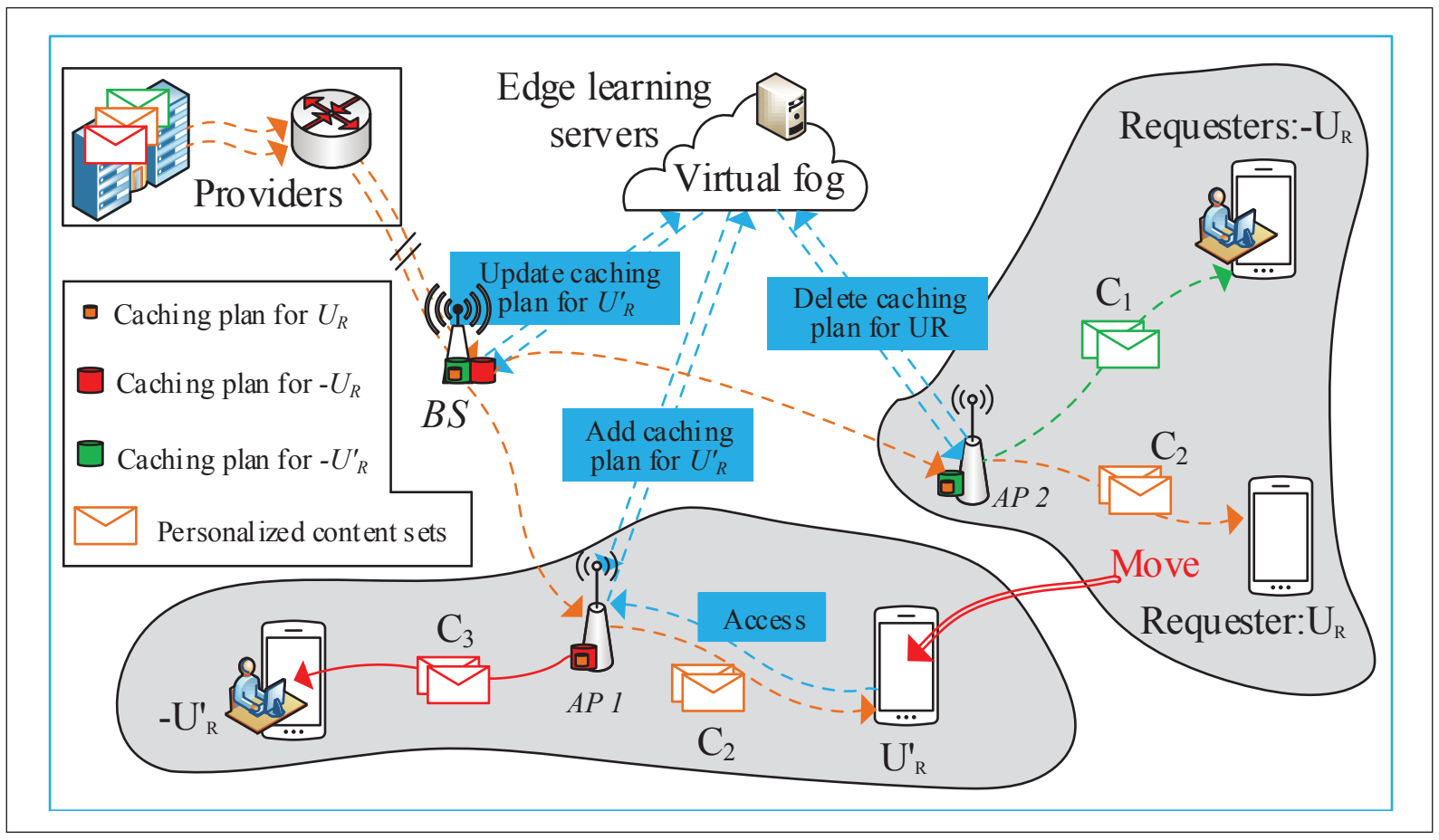

FIGURE 5. Mobile help pre-calling based on EdgeLaas in KCCH.

Different from the traditional redirection technology that needs global connectivity additional interaction, the mobile help pre-calling in KCCH allows upstream KCCH nodes to reserve a helper. For example, we can cache the helper's information on the base station when $U_R$ moves into a new access domain and updates its label as $U_R'$. $U_R'$ can gain the helper's information from the base station but not AP2.

On the basis of the above mentioned cases, EdgeLaaS can also be integrated into many other services (e.g., privacy protection) and important functions (e.g. knowledge traffic engineering.

## EdgeLaaS's Risks Control

While EdgeLaaS presents many disruptive promotions on connected healthcare, such orchestration over the underlying heterogeneous infrastructures still needs to pay attention to the following aspect's risks:

- Unbalanced resources distribution. There are many different types of entities (including terminals, routers, APs, and base stations) in connected healthcare systems. Different types of KCCH entities have different network resources. The network resources of these KCCH entities should be well allocated. Otherwise, it leads to non-robust EdgeLaaS.
- Untrusted knowledge access. The network of KCCH delivers knowledge retrieval and distribution rather than content interests and data packets. The decentralized knowledge sharing between KCCH nodes should be securely authenticated to circumvent the untrusted knowledge access.
- Appropriateness of the learning model. A bad learning model for EdgeLaaS will face many security problems (e.g., knowledge poisoning and adversarial examples).

Faced with these specified risks, we can use the following prevention methods to protect EdgeLaaS. First, the deployment of EdgeLaaS relies on context-aware dynamic resource offloading, which guarantees sufficient resources for edge nodes. Second, all knowledge exchange must pass through strict authentication and deep traffic inspection. Third, heterogeneous data is cooperatively analyzed in EdgeLaaS, and the medical knowledge is updated dynamically to circumvent adversarial attacks. Since humans have not cleared up the cause of adversarial examples, we suggest enhancing the security authentication to knowledge providers.

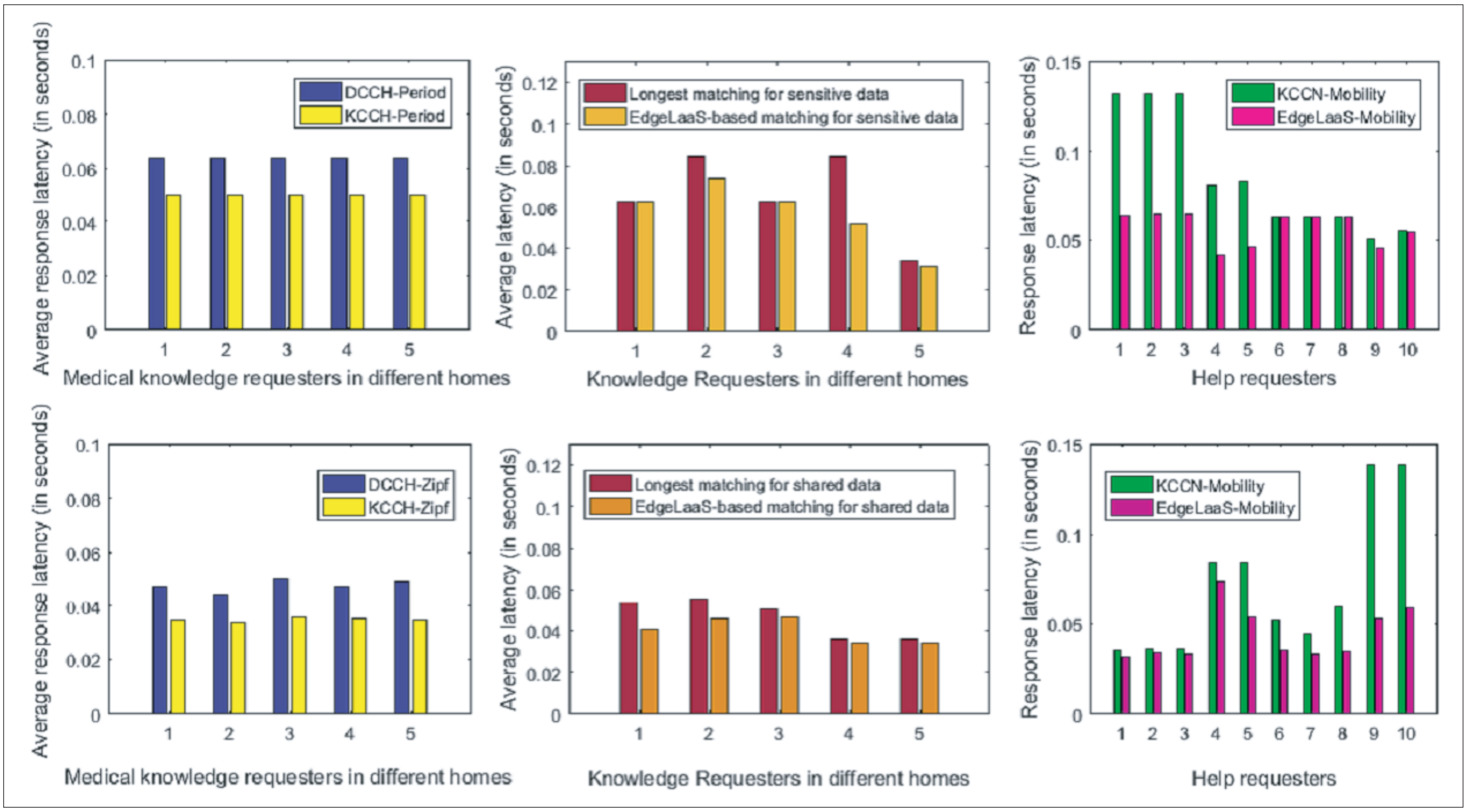

FIGURE 6. Performance evaluation of EdgeLaaS for KCCH.

## Performance Evaluation

In this section, we discuss the experimental results to validate the feasibility of EdgeLaaS for KCCH. In the experiments, we consider the average latency of knowledge retrieval under three different application scenarios as follows. To reduce the cost of artificial intelligence (AI) training, we can train some complicated models at the cloud and then deploy the pre-trained model at edge IoT nodes of healthcare. The results show that KCCH and EdgeLaaS can decrease the knowledge retrieval delay significantly.

For the first scenario, we consider a connected healthcare system with five homes. Each home contains two kinds of knowledge requesters: one is configured with the period request model, and the other is configured with the Zipf request model. Meanwhile, both cloud and fog are used to connect every home. There are one AP, one base station, one gateway, and the public Internet between a knowledge requester and the cloud data center, while the fog node provides knowledge directly through the base station. The left subplots of Fig. 6 show the average response delay of users' requests under DCCH and KCCH. As placing learning tasks on local fogs reduces the hop count between knowledge requesters and providers, it can be observed that the average latency of both KCCH-Zipf and KCCH-Period are less than DCCH-Zipf and DCCH-Period.

Different from the first scenario, all of the base stations in KCCH are enforced to join the fog networks when the second scenario works. Specifically, the knowledge requesters are divided into two different types according to the knowledge names that they are requesting. The learning applications deployed on fog nodes will aggregate the knowledge statistics and then adaptively provide fast self-help service for each requester. We also configure the requests generated by different requesters with the period request model and the Zipf request model, respectively. Influences of EdgeLaaS on average latency of knowledge retrieval with different requesters are shown as illustrated in the middle subplots of Fig. 6. The average latency of knowledge retrieval is reduced significantly by using EdgeLaaS. With this feature, EdgeLaaS can be exploited by network operators of KCCH to enforce more healthcare services.

The third scenario focuses on the issue of mobile help pre-calling. Twenty help requesters call for help randomly in this scenario. Each help requester has two different kinds of communication links: RAN and D2D. We exploit EdgeLaaS to achieve smart communication link selection in mobile circumstances. As an example for illustration, we record the average response latency for each help requester in mobile circumstances. The performance improvement of EdgeLaaS on patient mobility management is shown in the right subplots of Fig. 6.

## Conclusion

In this article, we review the enabling technologies of connected healthcare and envision that knowledge-centric connected healthcare will replace data-centric connected healthcare in the future. For KCCH, we introduce the knowledge generating, storing, naming, and routing principles. We then present the edge learning as a service framework to provide better demand response for patients. Two application cases of EdgeLaaS are implemented:

1. Fast self-help
2. Mobile help pre-calling
To validate the feasibility of proposed framework, we compare the average latency of three scenarios:
1. DCCH and KCCH
2. EdgeLaaS and no EdgeLaaS
3. Static and mobile help requests
As an emerging technology, the proposed EdgeLaaS and KCCH have the prospective to be applied in disaster response.

## Acknowledgments

This work is sponsored by the National Natural Science Foundation of China (Grant Nos. 61572355, U1736115, and 61431008).

## Biographies

Gaolei Li [S'15] received his B.S. degree in electronic information engineering from Sichuan University, Chengdu, China, in 2015 andhe is pursuing a Ph.D. degree in the School of Electronic Information and Electrical Engineering, Shanghai Jiao Tong University, China. He participates in many national projects, including ones with the National Natural Science Foundation of China, National "973 Planning of the Ministry of Science and Technology, China, and others. He was a TPC member of IEEE GLOBECOM (2017 and 2018), the International Conference on the Internet of Things (2016, 2017, and 2018), and IEEE SmartCity 2018. He won the best conference paper from the IEEE CSIM Technical Committee in 2018. His research interests are focused on fog computing, IoT, privacy-preserving in healthcare, and more.

Guangquan Xu [M'15] (losin@tju.edu.cn) received his Ph.D. degree from Tianjin University, China, in March 2008. He is currently a professor in the Big Data School at Qingdao Huanghai University, China. He is also a professor at the Tianjin Key Laboratory of Advanced Networking (TANK), College of Intelligence and Computing, Tianjin University. He is a member of the CCF. His research interests include cyber security and trust management.

Arun Kumar Sangaiah received his Master of Engineering degree from Anna University and Ph.D. from VIT University, Vellore, India, in 2007 and 2014, respectively. He is currently a professor at the School of Computing Science and Engineering, VIT University. In 2016, he was a visiting professor with the School of Computer Engineering at Nanhai Dongruan Information Technology Institute in China for six months. In addition, he has been appointed as a visiting professor at Southwest Jiaotong University, Chengdu, Changsha University of Science and Technology, China, Dongguan University of Technology, Guangdong, and Hwa-Hsia University of Technology, Taiwan. Further, he has visited many research centers and universities in China, Japan, and South Korea for joint collaboration toward research projects and publications. His areas of research interest include e-learning, learning engineering, machine learning, software engineering, computational intelligence, IoT, wireless networks, bio-informatics, and embedded systems. His outstanding scientific production spans over 230+ contributions published in high standard ISI journals, such as *IEEE Communication Magazine*, *IEEE Systems*, and *IEEE IoT*. His publications are distributed as follows: 230 papers indexed and 21 papers indexed in Scopus. In addition, he has authored/edited 8 books (Elsevier, Springer and others) and edited 50 Special Issues in reputed ISI journals, such as *IEEE Communications Magazine*, *IEEE IoT*, and *ACM Transactions on Intelligent Systems and Technology*. He has also registered one Indian patent in the area of computational intelligence. His Google Scholar Citations reached 1300+ with h-index: 20 and i10-index: 39. Finally, he is an Editorial Board member and Associate Editor of many reputed ISI journals.

Jun Wu [S'08, M'12] (junwuhn@sjtu.edu.cn) is an associate professor of the School of Cyber Security, Shanghai Jiao Tong University, Shanghai Key Laboratory of Integrated Administration Technologies for Information Security, and vice director of the National Engineering Laboratory for Information Content Analysis Technology, China. He obtained his Ph.D. degree in information and telecommunication studies at Waseda University, Japan. He was a postdoctoral researcher for the Research Institute for Secure Systems (RISEC), National Institute of Advanced Industrial Science and Technology (AIST), Japan, from 2011 to 2012. He worked as a researcher for the Global Information and Telecommunication Institute (GITI), Waseda University from 2011 to 2013. He has hosted and participated in several research projects for the National Natural Science Foundation of China, National 863 Plan and 973 Plan, Japan Society of the Promotion of Science (JSPS) projects, and others. He is an Associate Editor of *IEEE Access*. He has been a Guest Editor for the *IEEE Sensors Journal* and a TPC member of more than 10 international conferences including EAI WINCON, IEEE ICC, IEEE GLOBECOM, and more. His research interests include advanced computing and communications techniques, smart grids, Internet of Things, industrial security, and so on.

Jianhua Li is a professor/Ph.D. supervisor and the dean of the School of Information Security Engineering, Shanghai Jiao Tong University, Shanghai Key Laboratory of Integrated Administration Technologies for Information Security, China. He got his B.S., M.S., and Ph.D. degrees from Shanghai Jiao Tong University in 1986, 1991, and 1998, respectively. He was the chief expert in the information security committee experts of the National High Technology Research and Development Program of China (863 Program). He is also a committee expert on the Information Technique Standardization Committee of Shanghai. He was the leader of more than 30 state/province projects of China, and published more than 300 papers. He authored over six books and holds about 30 patents. He made three standards and has five software copyrights. He is a member of the committee of information security area of the state 10th five-year plan of China. He got the First Prize of National Technology Progress Award of Shanghai in 2003 and 2004, and he got two First Prize of National Technology Progress Awards of Shanghai in 2004. He got the Second Prize of National Technology Progress Award of China in 2005. He was the Chief Expert in the Information Security Committee Experts of National High Technology Research and Development Program of China of China. His research interests include cyberspace security, data science, next-generation networks, smart grid, Industrial Internet of things (IIoT), and so on.